\documentclass[a4paper,11pt]{article}

\usepackage{tikz}
\usetikzlibrary{quantikz}
\usepackage{ifxetex,ifluatex}
\usepackage{geometry}
\usepackage{hyperref,graphicx,float,amsmath,amsfonts,amssymb,braket,tabularx,caption}

\if\ifxetex T\else\ifluatex T\else F\fi\fi T%
  \usepackage{fontspec}
\else
  \usepackage[T1]{fontenc}
  \usepackage[utf8]{inputenc}
  \usepackage{lmodern}
  \usepackage{url}
\fi
\usepackage{authblk}
\usepackage{braket}
\usepackage{cite}
\usepackage{tikz}
\usepackage{caption}
\usepackage{subcaption}
\usepackage[bottom]{footmisc}
\usepackage{multirow}
\usepackage{resizegather}
\usepackage{hyperref}
\newsavebox{\imagebox}

\title{Towards optimization under uncertainty for fundamental models in energy markets using quantum computers}
\author{M.C. Braun}
\author{T. Decker}
\author{N. Hegemann}
\author{S.F. Kerstan}
\author{F. Lorenz}
\affil{JoS QUANTUM GmbH, Frankfurt am Main, Germany \footnote{For contact email: niklas.hegemann@jos-quantum.de. Authors are listed in alphabetical order.}}

\begin{document}
\maketitle
\vspace{-1cm}
\abstract{We present a method to formulate the unit commitment problem in energy production as
quadratic unconstrained binary optimization (QUBO) problem,
which can be solved by classical algorithms and quantum computers. We suggest a first approach to consider uncertainties in the renewable energy supply, power demand and machine failures. We show how to find cost-saving solutions of the UCP under these uncertainties on quantum computers. We also conduct a study with different problem sizes and we compare results of simulated annealing with results from quantum annealing machines. }

\section{Introduction}
Electrical energy production and supply is a foundation of developed economies and essential for the stability of societies. Energy markets provide producers, consumers and grid operators an efficient way for pricing electricity, depending on forecasted and actual demand and supply. As the grid frequency needs to be stable (50 Hz in Europe) and as big amounts of electrical energy cannot be stored efficiently, transmission operators need to balance the output of power generation units and the consumption in real-time.

Energy production based on solar and wind power depends on weather conditions
and these can change rather quickly, introducing extreme volatility in production levels and prices. The European Union plans to continue the substitution of power plants powered by fossil fuels, i.e. coal, oil and gas, with renewable sources of energy \cite{REPowerEU}.
This introduces an even higher volatility while forecasting the feed-in to the energy grid is limited. Therefore, smart ways to steer the production as well as the consumption need to be introduced to produce energy with the lowest cost while satisfying demand. Finding a power plant dispatch schedule whose expected supply matches the demand at any given point in time and comes with the lowest possible costs is a well-known optimization problem in energy markets.

Energy prices are usually derived from the power plant with the lowest marginal costs that match the energy demand. This is called merit order \cite{meritorder}. Beside this, a number of technical constraints on the power generating units must be observed. Conventional power plants cannot freely be switched on and off and have to obey limits in their power generation range. The problem of finding a suitable power unit schedule is known as unit commitment problem (UCP) and it is NP-hard to solve. Several approaches exist to solve such types of problems including dynamic programming \cite{dynprogramming}, Lagrangian relaxation \cite{lagrangian}, Benders decomposition \cite{bender}, mixed integer programming \cite{MIP} and reinforcement learning \cite{RL}.

The feed-in from renewable power plants like solar parks and wind farms depend on factors like sunshine hours and wind speed. The accuracy of predictions of the weather is limited and this leads to an unknown residual future power demand that conventional power plants have to compensate. Other uncertainties come in the form of machine failure events or unpredictable demand fluctuations on small time scales. Considering such uncertainties in the UCP leads to the stochastic unit commitment problem (SUCP). There exist many approaches to the SUCP and an overview is given in~\cite{Haberg}. Since it can be shown that the UCP is NP-hard~\cite{ucp} (and so is the SUCP), there exists no efficient method to solve this problem exactly. Instead, approximate solutions and solutions to relaxed problems are used in practice.

Quantum computing may open new ways of solving the UCP and  the SUCP efficiently.
Recent theoretical work has shown that with quantum algorithms we can achieve advantages over classical methods for several problems in theory~\cite{Montanaro, Chakrabarti}. One way is to use annealing methods that are carried out by Ising machines. One example of commercially available quantum annealers are the machines from D-Wave. These machines leverage quantum mechanical effects like quantum tunneling to find optimal solutions to annealing problems quickly. This opens the path to provide solutions with higher quality in less time.

Although the technology still has to overcome technical limitations before reliable large-scale quantum computers are available, the fast progress in the field promises a true alternative to classical solvers in the future. In this paper, we explore the possibility of formulating the UCP in terms of a binary quadratic unconstrained optimization (QUBO) problem that can be solved by many different solvers, e.g. by quantum inspired algorithms \cite{SBM}, quantum annealers \cite{QA,perspectiveQA} and variational quantum algorithms like the variational quantum eigensolver \cite{vqe} or the quantum approximate optimization algorithm \cite{qaoa}. We use simulated annealing \cite{SA} as well as quantum annealing to find solutions for our QUBO formulations. 

A similar approach for the static UCP was published recently in~\cite{Fraunhofer}. Here, we go beyond this method by advancing the QUBO formulation to solve a relaxed version of the UCP. In this version, we assume that the overall supply needs to match the overall demand, but at each timestep supply and demand can deviate and a small over- or undersupply can be compensated with storage devices.

The results for several examples  were obtained by classical solver algorithms and by D-Wave annealing hardware. This work might help hardware companies to better understand the hardware requirements for real-world business applications.

\section{Elements of the unit commitment problem} \label{ucp}
In this work, we restrict ourselves to the single bus approximation where the geographical structure of the electricity grid is ignored. The demand is considered to be satisfied whenever the total production equals the total demand. The UCP refers to the optimization task of finding the cost optimized production plan of $N$ conventional power generators whose total energy supply adds up to ${\rm demand}(t)$ at every timestep $t \in \{1,2,\dots,T\}$.
Each power generator $k \in \{1,2,\dots,N\}$ can generate a power output ${\rm gen}_k$ with variable costs ${\rm varcost}_k$, i.e. the costs per produced unit in a timestep. Startup costs ${\rm startcost}_k$ are introduced whenever a unit is turned on.

We extend the problem by introducing $R$ renewable units that run without variable costs and that have a fixed power-time profile ${\rm supply}^{\rm RE}_r(t)$ which is determined by external factors, like the weather or tides. These renewable units effectively reduce the required amount of energy and give a residual demand that has to be compensated by conventional power plants. We call this an effective demand $d(t)$ and we require a solution of the UCP to provide
\begin{equation*}
    d(t)= \rm{demand}(t)-\sum_{r=1}^R supply^{RE}_r(t) 
    =\sum_{k=1}^N gen_k(t) \;\; {\rm for\; all} \;\; t \in \{1,2,\dots,T\}~.
\end{equation*}

More complexity arises when restrictions that originate from power generators are introduced. The power generating units need to obey the following constraints: The power supply of each unit has a lower bound ${\rm mingen}_k$ and an upper bound ${\rm maxgen}_k$, i.e., ${\rm gen}_k(t) \in [{\rm mingen}_k,{\rm maxgen}_k]$ for all $t \in \{1,2,\dots,T\}$.
Furthermore, power generators exhibit technical minimum run times ${\rm minup}_k$ and minimum shut down times ${\rm mindown}_k$. This means that once power unit $k$ is turned on it can not be turned off after running for at least ${\rm minup}_k$ timesteps. A similar condition ensures compliance of downtimes.
To summarize, a specific problem instance of the UCP is fully determined by the following set of parameters:

\begin{itemize}
\item ${\rm demand}(t)$ for $t\in\{1,\ldots,T\}$
\item ${\rm supply}^{\rm RE}_r(t)$ for $r \in \{1,\ldots, R\}$ and $t\in\{1,\ldots,T\}$
\item ${\rm mingen}_k, {\rm maxgen}_k, {\rm varcost}_k, {\rm minup}_k, {\rm mindown}_k$ for $k \in \{1, \ldots, N\}$
\end{itemize}

\section{QUBO formulation with penalty terms} \label{QUBO}
\subsection{Formulation of QUBO problems}
In this section, we discuss techniques that can be used to formulate combinatorial optimization problems as Quadratic Unconstrained Binary Optimization (QUBO) problems. It is a mathematical construct equivalent to Ising spin glasses. Once formulated, these QUBO problems can be solved by different techniques, including quantum annealers, quantum computers and other types of Ising-machines~\cite{lucas}.

Such optimization problems can be formulated as follows when we set $\mathbb{B}=\{0,1\}$.
Let $f_Q:\mathbb{B}^n \rightarrow \mathbb{R}$ be the quadratic polynomial
\begin{equation}
f_Q(x_1,\ldots, x_n) = \sum^n_{i=1} \sum^n_{j=1} q_{ij} x_i x_j  \label{qubo1}
\end{equation}
with $q_{ij} \in \mathbb{R}$, which can be understood as real valued entries of a matrix $Q$. The function $f_Q$  is called the objective function or objective of the corresponding QUBO matrix $Q$. For properly chosen coefficients $q_{ij}$, the global extrema of the objective function correspond to a solution of an optimization problem, where the solution itself is encoded in the binary-valued vector $x=(x_1,\ldots,x_n) \in \mathbb{B}^n$ that fulfills\footnote{In the following, we write vectors $(x_1,\ldots,x_n)^T$ as row vectors $(x_1,\ldots,x_n)$ to simplify notation and to avoid ambiguities with the number $T$ of timesteps.}
\begin{equation*}
\min(f_Q) = \sum^n_{i=1} \sum^n_{j=1} q_{ij} x_{i} x_{j} \quad {\rm or} \quad
\max(f_Q) = \sum^n_{i=1} \sum^n_{j=1} q_{ij} x_{i} x_{j}\,.
\end{equation*}

A more elegant way to write equation~\eqref{qubo1} is using Dirac's bracket notation:
\begin{equation*}
f_Q(x) = \Braket{x|Qx}
\end{equation*}
Here, $\langle x|y\rangle$ is the scalar product of two vectors $\ket{x}$ and $\ket{y}$. The matrix $Q$ acts as an operator on the corresponding vector space and $\bra{x}$ is the adjoint of $\ket{x}$. Subsequently, this notation is used in some cases for simplifying the notation. In analogy to the quantum mechanical origin of this notation, the solution to a minimization problem corresponds to a global minimum in the energy landscape.

An efficient QUBO solver algorithm is a machinery that takes $Q$ as an input and  outputs the vector $|x\rangle$ of the global minima in a reasonable amount of time, i.e. such an algorithm would provide us with a tool to solve optimization problems of this kind. To the best of our knowledge, there exists no reliable and fast algorithm to solve universal problems of this type as it can be shown that QUBO problems are NP-hard \cite{glover}. However, several kinds of algorithms exist that can be used to find approximate good solutions. Here, we compare simulated annealing to the results of hardware-based quantum annealing. A quantum annealer represents a special type of quantum computer that performs approximate adiabatic quantum annealing and is able to provide a QUBO solver, at least on a theoretical level.

\subsection{Building the QUBO for the UCP}\label{buildqubo}
The strategy for defining the QUBO elements $q_{ij}$ is to use quadratic terms that give higher objective values for expensive and forbidden states and lower objective values for cheaper and allowed states, respectively. Forbidden and allowed in this context refers to the adherence to constraints. As introduced in section~\ref{ucp}, we want the following conditions and technical constraints to be fulfilled in the UCP:
\begin{itemize}
    \item minimizing costs
    \item matching power demand
    \item obey minimum and maximum power generation constraints
    \item obey minup and mindown times
\end{itemize}
We start the QUBO formulation of the UCP by introducing a resolution of the power supply in the form of a binary representation of numbers. Since we work in a discrete model, it is not possible for ${\rm gen}_k(t)$ to take continuous values between ${\rm mingen}_k$ and ${\rm maxgen}_k$. However, we can approximate the continuous case arbitrarily well by introducing a discretization of the interval $[{\rm mingen}_k,{\rm maxgen}_k]$ and we write
\begin{equation}\label{eq:powerdiscrete}
    {\rm maxgen}_k-{\rm mingen}_k=\sum_{b=1}^B p_{k,b}
    \;{\rm with}\; p_{k,b}=\frac{2^{b-1} \cdot (\rm{maxgen}_k-\rm{mingen}_k)}{2^B-1}.
\end{equation}
Here, the number of bits used for the discretization is $B$.
 Furthermore, we introduce the variables $x^p_{k,b}(t)$ for $b \in \{1,2,\dots,B\}$
for each timestep $t$. The variable $x^p_{k,b}(t)$ describes how much the power supply of unit $k$   contributes to the sum \eqref{eq:powerdiscrete} depending on level $b$. We also introduce the binary variables $x_k^{1/0}(t)$, which describe whether unit $k$ is on or off in timestep $t$.
Then we can write
\begin{equation}\label{gen}
    {\rm gen}_k(t)={\rm mingen}_k \cdot x_k^{1/0}(t) + \sum_{b=1}^B p_{k,b} \cdot x^p_{k,b}(t) = {\rm mingen}_k \cdot x_k^{1/0}(t) + \langle p_k|x_k^p(t)\rangle
\end{equation}
with $p_k=(p_{k,1},\ldots,p_{k,B})$ and $x_k^p(t)=(x_{k,1}^p(t),\ldots, x_{k,B}^p(t))$,
which suits the QUBO formalism. Increasing the resolution $B$ increases the number $R(B)=2^B$ of possible power supply realizations and it improves the fineness 
$$
F(B) = \frac{{\rm maxgen}_k-{\rm mingen}_k}{2^B-1}
$$ 
of the power range as can be seen in table~\ref{tab:resolutionoverhead}.
The introduction of an additional variable $x_k^s(t)$ per unit and timestep is required to describe the start of the power production of a unit. If unit $k$ is off in timestep $t-1$ and if it is turned on in timestep $t$, then we set the start-variable $x_k^s(t)=1$ and $x_k^s(t)=0$ else.
\begin{table}[h]
\captionsetup{width=0.87\linewidth,font={small},format=hang}
\centering
\caption{Discretization of the interval $[x,y]$. Number $R(B)$ of power supply realizations and the relative fineness $F(B)/(y-x)$ as functions of the resolution parameter $B$.}
    \begin{tabular}{r|r|l}
        $B$ & $R(B)$ & $F(B)/(y-x)$ \\
        \hline 
         2  &   4  &   0.3333...  \\
         3  &   8  &   0.1429...   \\
         4  &   16  &   0.0607... \\
         6  &   64  &   0.0159... \\
         10  &   1024  &   0.001... \\
         20 & 1048576 & $ \approx 1\cdot 10^{-6}$ \\
    \end{tabular}
    \label{tab:resolutionoverhead}
\end{table}

For timestep $t$, we choose the corresponding part of the full solution vector to be arranged like
\begin{equation*} 
    x(t)=(x^p(t),x^{1/0}(t),x^s(t)),
\;{\rm where}\;\;
x^p(t)=(x^p_{1}(t),x^p_{2}(t),\dots,x^p_{N}(t))
\end{equation*}
encodes the power supply stages of all units. The vectors
$$
x^{1/0}(t)=(x_1^{1/0}(t),x_2^{1/0}(t),\dots,x_N^{1/0}(t)) \quad {\rm and} \quad x^{s}(t)=(x_1^{s}(t),x_2^{s}(t),\dots,x_N^{s}(t))
$$
contain the on/off and start information, respectively. Each element $x^p_k(t)$ has the structure 
$$
x^p_k(t)=(x^p_{k,1}(t),x^p_{k,2}(t),\dots,x^p_{k,B}(t)).
$$
The introduction of $x^{1/0}(t)$ enables us to implement the minimum power constraint, which comes at the cost of increasing the problem size. The full solution vector for $T$ timesteps is given by 
\begin{equation}\label{eq:solvector}
    x=(x(1),x(2),\dots,x(T)).
\end{equation}
The length of this vector is given by ${\rm dim}(x)=TN(B+2)$, i.e. it grows linearly with the number of timesteps $T$, units $N$ and resolution $B$. 
The next step is to build the QUBO matrix $Q$ by translating the UCP into matrix elements. We do this step by step and implement the individual problem specification via defining corresponding penalty terms, which are added together at the end. 

The minimal costs for the optimal power unit allocation include the variable costs, i.e. the running cost per unit time, as well as start costs. A suitable penalty term is given by
\begin{equation}\label{lincostterm}
    P^{\rm cost} \cdot \sum_{t=1}^T \sum_{k=1}^N {\rm varcost}_k \cdot \left( {\rm mingen}_k \cdot x_k^{1/0}(t) + \Braket{p_k|x_k^p(t) }  \right)+ {\rm startcost}_k \cdot x_k^s(t),
\end{equation}
where $P^{\rm cost}$ is a penalty parameter that has to be adjusted. Note that this form is equivalent to equation~\eqref{qubo1}, since for binary variables we can use the equality $x=x^2$ to substitute linear terms with quadratic ones. 
It is also possible to directly introduce a quadratic cost term by squaring~\eqref{lincostterm}. We obtain
\begin{equation*}
    P^{\rm cost} \cdot \left( \sum_{t=1}^T \sum_{k=1}^N {\rm varcost}_k \cdot \left( {\rm mingen}_k \cdot x_k^{1/0}(t) + \Braket{p_k|x_k^p(t) }\right) + {\rm startcost}_k \cdot x_k^s(t) \right)^2
\end{equation*}
and this enlarges the differences of the objective function for different states by increasing the range of the energy spectrum. We quantify this by defining
\begin{equation*}
    \kappa_k(x(t))={\rm varcost}_k \cdot \left( {\rm mingen}_k \cdot x_k^{1/0}(t) + \Braket{p_k|x_k^p(t) } \right)+ {\rm startcost}_k \cdot x_k^s(t) 
\end{equation*}
and observe that $\kappa_k(x(t)) \in [0,\kappa_k^{\rm max} ]$ where $\kappa_k^{\rm max}={\rm varcost}_k \cdot {\rm maxgen}_k + {\rm startcost}_k$ is the upper bound. It is now obvious that 
\begin{equation*}
P^{\rm cost} \cdot \sum_{t=1}^T \sum_{k=1}^N \kappa_k(x(t)) < P^{\rm cost} \left( \sum_{t=1}^T \sum_{k=1}^N \kappa_k(x(t)) \right)^2 
\end{equation*}
if $\sum_{t=1}^T \sum_{k=1}^N \kappa_k(x(t)) > 1$. Since the penalty term can only take non-negative values, we can rescale the costs to ensure that this is always true. Hence, the energy spectrum in the case of the quadratic term has a larger range. Also, if we additionally define $\kappa^{\rm max}=\max \{\kappa^{\rm max}_1,\kappa^{\rm max}_2,\dots,\kappa^{\rm max}_K \}$, the following inequalities are true:
\begin{subequations}
    \begin{align*}
        P^{\rm cost} \cdot \sum_{t=1}^T \sum_{k=1}^N \kappa_k(x(t)) &< P^{\rm cost} TN\kappa^{\rm max}\\
        P^{\rm cost} \cdot \left( \sum_{t=1}^T \sum_{k=1}^N \kappa_k(x(t))\right)^2 &< P^{\rm cost} T^2N^2 (\kappa^{\rm max})^2
    \end{align*}
\end{subequations}
This shows that the energy range is potentially larger in the case of a quadratic cost term. From now on, we choose to continue with the quadratic cost term.  \\

The next step is to match the power demand in every timestep. We require the supply resulting from an optimal power unit schedule to match the forecasted energy amount for all timesteps. As described in section \ref{ucp}, we use an effective demand $d(t)$ in every timestep and find that a suitable penalty term is given by
\begin{equation}\label{demandterm}
    P^{\rm demand} \cdot \sum_{t=1}^{T}  \left( \sum_{k=1}^N{\rm  mingen}_k \cdot x_k^{1/0}(t) + \Braket{p_k|x_k^p(t)} - d(t) \right)^2 ~.
\end{equation}
Again, $P^{\rm demand}$ is a penalty strength which has to be chosen properly.
Note that this gives multiple terms of the form $d(t)^2$, which do not include $x$ and therefore cannot take the form of equation~\eqref{qubo1}. However, we can subtract these constant terms from~\eqref{demandterm} and this leads to a constant energy shift for all states. Consequently, minimizing this new function also minimizes \eqref{demandterm}. \\

We now add a penalty to respect the technical restriction of minimum run times. We have specified the minimum run time of unit $k$ via ${\rm minup}_k$. When unit $k$ is turned on in timestep $t$, the earliest shutdown of this unit is possible after timestep $t+{\rm minup}_k$, i.e. the unit has to run for ${\rm minup}_k$ timesteps. With the penalty strength $P^{\rm minup}$, we use the penalty term
\begin{equation*}
P^{\rm minup} \cdot   \sum_{t=1}^{T}  \sum_{k=1}^N  x_k^s(t) \bigg( {\rm minup}_k \cdot x_k^s(t)-\sum_{\tau=t}^{t+{\rm minup}_k-1} x_k^{1/0}(\tau) \bigg)~.
\end{equation*}
If the minimum run time condition is fulfilled, the penalty vanishes. Note that in the summation over $\tau$ we only consider elements up to at most $\tau=T$.\\

The penalty terms for minimum down times work in a similar way. When unit $k$ is turned off in timestep $t$, the earliest possible start of this unit is after timestep $t+{\rm mindown}_k$, i.e., the unit has to be shut down for ${\rm mindown}_k$ timesteps. Introducing the penalty strength $P^{\rm mindown}$, we choose
\begin{equation*}
P^{\rm mindown} \cdot \sum_{t=1}^{T}  \sum_{k=1}^N \left( x^{1/0}_k(t-1)-x^{1/0}_k(t) +x_k^s(t)\right) \cdot \sum_{\tau=t}^{t+{\rm mindown}_k-1} x^{1/0}_k (\tau)
\end{equation*}
as the corresponding term in the QUBO. As in the case of minimum run time, fulfilling the minimum down time condition gives a zero penalty. Note that we do not sum over elements for values of $t-1$ and $\tau$ that are outside of $\{1,\ldots,T\}$.

The minimum and maximum power supply of each power generator are automatically met due to equation~\eqref{gen}, if we can ensure that whenever at least one of the $x_{k,b}^p(t)$ is non-zero, the variable $x^{1/0}_k(t)$ is also set to one. This requires a penalty term of the form 
\begin{equation*}
    P^{{\rm inter}_1} \cdot \sum_{t=1}^{T}  \sum_{k=1}^N    \sum_{b=1}^B x^p_{k,b}(t) \left( 1-x^{1/0}_k(t) \right),
\end{equation*}
where $P^{{\rm inter}_1}$ denotes the penalty strength. This term interrelates the variables $x_{k,b}^p(t)$ and $x^{1/0}_k(t)$. \\

Finally, setting the start variable is required to ensure that we correctly connect the start variable $x_k^s(t)$ to the on/off-variable $x_k^{1/0}(t)$ of unit $k$ such that we have
\begin{equation*}
x_k^s(t)=
    \begin{cases}
        1 & \text{for } x_k^{1/0}(t-1)=0 \;{\rm and}\; x_k^{1/0}(t)=1 \\
        0 & \text{else}
    \end{cases}
\end{equation*}
A suitable penalty term is  
\begin{equation*}
    P^{{\rm inter}_2}  \cdot \sum_{t=1}^{T}  \sum_{k=1}^N \bigg( \left( x_k^{1/0}(t+1) - x^s_k(t+1) \right)^2 + x^{1/0}_k(t) \cdot  \left(x^s_k(t+1) - x^{1/0}_k(t+1) \right) \bigg)~
\end{equation*}
with the corresponding penalty strength $P^{{\rm inter}_2}$. Here, we do not sum over elements for values of $t+1$ and $\tau$ that are outside of $\{1,\ldots,T\}$.\\

We can now write the full QUBO as the sum of the penalty terms above:

\begin{align}\label{eq:fullQUBO}
    \begin{split}
    Q(x)=& P^{\rm cost} \left( \sum_{t=1}^T \sum_{k=1}^N {\rm varcost}_k \cdot \left( {\rm mingen}_k \cdot x_k^{1/0}(t) + \Braket{p_k|x_k^p(t) } \right)+ {\rm startcost}_k \cdot x_k^s(t) \right)^2 \\
    +& P^{\rm demand} \cdot \sum_{t=1}^{T}  \left( \sum_{k=1}^N {\rm mingen}_k \cdot x_k^{1/0}(t) + \Braket{p_k|x_k^p(t)} - d(t) \right)^2 \\
    +& P^{\rm minup} \cdot   \sum_{t=1}^{T}  \sum_{k=1}^N  x_k^s(t) \bigg( {\rm minup}_k \cdot x_k^s(t)-\sum_{\tau=t}^{t+{\rm minup}_k-1} x_k^{1/0}(\tau) \bigg) \\
    +& P^{\rm mindown} \cdot \sum_{t=1}^{T}  \sum_{k=1}^N \left( x^{1/0}_k(t-1)-x^{1/0}_k(t)+x_k^s(t)\right) \cdot \sum_{\tau=t}^{t+{\rm mindown}_k-1} x^{1/0}_k (\tau) \\
    +&P^{{\rm inter}_1} \cdot \sum_{t=1}^{T}  \sum_{k=1}^N   \sum_{b=1}^B x^p_{k,b}(t) \left( 1-x^{1/0}_k(t) \right) \\
    +& P^{{\rm inter}_2}  \cdot \sum_{t=1}^{T}  \sum_{k=1}^N \bigg( \left( x_k^{1/0}(t+1) - x^s_k(t+1) \right)^2 + x^{1/0}_k(t) \cdot  \left(x^s_k(t+1) - x^{1/0}_k(t+1) \right) \bigg)
    \end{split}
\end{align}

\subsection{Tuning penalty parameters}
In this section we derive relations between the different penalty strength parameters that appear in equation~\eqref{eq:fullQUBO} to ensure that the model correctly prioritizes different problem constraints and relates good solutions of the UCP to low energy states. 

The first requirement is that we want to avoid demand mismatches in favor of cost savings. Therefore, consider a solution $x$ that matches the demand for all timesteps. Further, a second solution is given which equals the first one except that in time $t'$ there is a supply deficit of $\Delta {\rm gen}(t')={\rm gen}(t')-d(t')<0$ which comes at the cost saving $\Delta \kappa(t')<0$. Using equation~\eqref{eq:fullQUBO}
and the Kronecker delta notation $\delta_{t,t'}$, the corresponding condition for the difference of these two solutions is
\begin{align*}
\begin{split}
P^{\rm cost}\cdot \left( \sum_{t=1}^T \sum_{k=1}^N \kappa_k(x(t))\right)^2
&<
P^{\rm cost}\cdot \left( \sum_{t=1}^T \left[\, \sum_{k=1}^N  \kappa_k(x(t))+ \Delta \kappa(t') \delta_{t,t'} \right] \right)^2 \\
&+ P^{\rm demand} \Delta {\rm gen}(t')^2
\end{split}
\end{align*}
We now exploit the following inequality: With $\epsilon<0<|\epsilon|<z$ it follows that  $\left( z+\epsilon \right)^2<(z+\epsilon)(z-\epsilon)=z^2-\epsilon^2$. We use this to estimate the right hand side and with a multiplication with $N^2$ and we obtain:
\begin{align*}
\begin{split}
P^{\rm cost}\cdot \left( \sum_{t=1}^T \sum_{k=1}^N \kappa_k(x(t))\right)^2 <& P^{\rm cost}\cdot \left( \sum_{t=1}^T \sum_{k=1}^N \kappa_k(x(t))\right)^2 - P^{\rm cost} N^2 \Delta \kappa(t')^2 \\
&+ P^{\rm demand} \Delta {\rm gen}(t')^2~.
\end{split}
\end{align*}
This inequality holds for all possible supply deficits in one timestep and all cost savings. To get the best bound, we  take the smallest possible demand deficit $\min_{k,b}(p_{k,b})$ and the largest possible value
$$
\Delta \widehat{\kappa}= \max_k{({\rm varcost}_k \cdot [{\rm maxgen}_k-{\rm mingen}_k]+{\rm startcost}_k)}
$$ 
for cost saving. These values yield
\begin{equation*}
    P^{\rm demand}>\frac{N^2 \Delta \widehat{\kappa}^2}{\min_{k,b}{(p_{k,b})}^2}P^{\rm cost}~.
\end{equation*}
To fullfill this inequality in all cases, we choose 
\begin{equation*}
    P^{\rm demand}= 2 \cdot \frac{N^2 \Delta \widehat{\kappa}^2}{\min_{k,b}{(p_{k,b})}^2}P^{\rm cost}~.
\end{equation*}
Since the penalty parameter $P^{\rm cost}$ defines the energy scale of the objective function,  we can set $P^{\rm cost} = 1$ without loss of generality.

The other penalty parameters represent hard constraints of the problem and  have to be prioritized over the cost and demand matching, i.e., we want to penalize violations of these technical constraints more than demand mismatches or high costs. Empirically, we find that suitable values for these parameters are given by
\begin{subequations}
\begin{align*}
    P^{\rm minup}&=P^{\rm mindown}=10^2 \cdot N^2 \Delta \widehat{\kappa}^2~, \\
    P^{{\rm inter}_1}&= 10^4 \cdot N^2 \Delta \widehat{\kappa}^2\;\;{\rm and} \\
    P^{{\rm inter}_2}&=10^6 \cdot N^2 \Delta \widehat{\kappa}^2~.
\end{align*}
\end{subequations}

\subsection{Example}\label{example1}
We now present results of the method applied to a small-sized problem, which consists of $N=2$ power supply units and $T=3$ timesteps with a resolution of $B=10$. We also consider two renewable supply units, which contribute fixed, but time dependent amounts of energy. As a result, the two power units have to match the residual demand. The individual parameters of each unit are given in table~\ref{tab:unitparameters1}. 

\begin{table}[h]
    \captionsetup{width=0.87\linewidth,font={small},format=hang}
    \centering
    \begin{tabular}{c|c|c|c|c|c|c}
        unit $k$ & ${\rm varcost}_k$ & ${\rm startcost}_k$ & ${\rm mingen}_k$ & ${\rm maxgen}_k$  & ${\rm minup}_k$ & ${\rm mindown}_k$ \\
        \hline
         1 & 65 &200 & 34 & 505 & 2 & 1 \\
         2 & 25 &500 & 250 & 900 & 2 & 2 
    \end{tabular}
    \caption{Parameters of the power generating units for the example.}
    \label{tab:unitparameters1}
\end{table}

Table~\ref{tab:demand1} shows the power demand as well as the supply from renewables for three timesteps and the resulting effective demand.
\begin{table}[H]
    \captionsetup{width=0.87\linewidth,font={small},format=hang}
    \centering
    \begin{tabular}{c|c|c|c|c}
         $t$ & ${\rm demand}(t)$ & ${\rm supply}^{\rm RE}_1(t)$ & ${\rm supply}^{\rm RE}_2(t)$ & $d(t)$ \\
         \hline
          1 & 618 & 50 & 100 & 468\\
          2 & 1145 & 50 & 150 & 945\\
          3 & 710 & 25 & 125 & 560
    \end{tabular}
    \caption{Power demand, renewable supply and effective demand for the individual timesteps.}
    \label{tab:demand1}
\end{table}

\begin{figure}[H]
\captionsetup{width=0.9\linewidth,font={small},format=hang}
    \centering
    \includegraphics[width=0.65\textwidth]{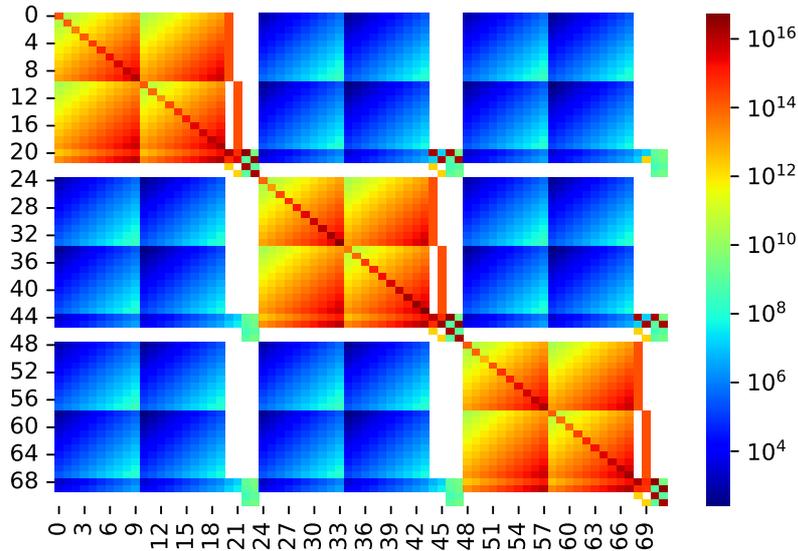}
    \caption{Heatmap of the $(72 \times 72)$-dimensional QUBO matrix for the UCP instance specified by the parameters from table~\ref{tab:unitparameters1} and table~\ref{tab:demand1}.}
    \label{fig:qubo1}
\end{figure}

\begin{figure}[H]
    \captionsetup{width=0.87\linewidth,font={small},format=hang}
     \centering
     \begin{subfigure}[t]{0.49\textwidth}
         \centering
         \includegraphics[width=\textwidth]{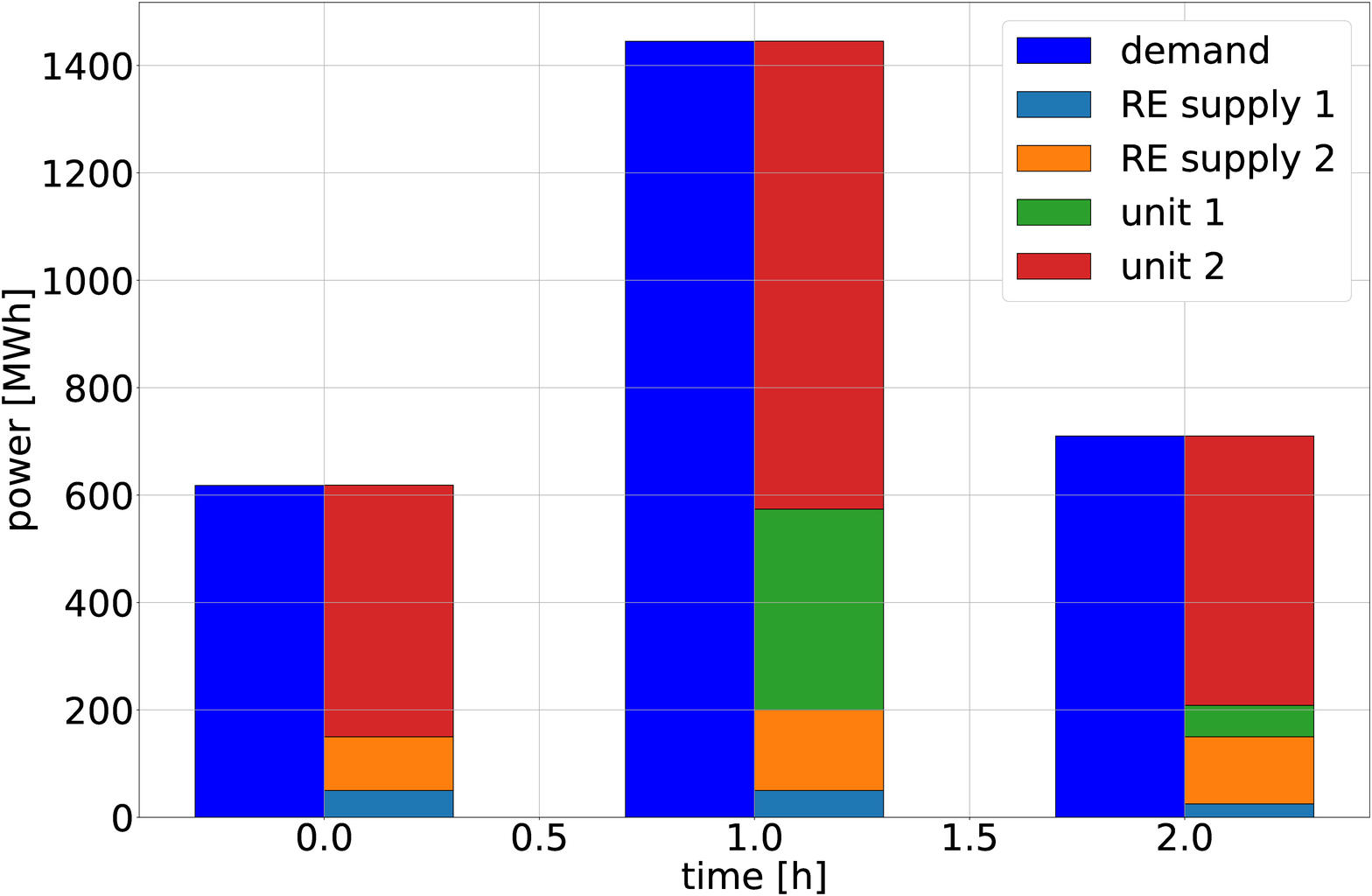}
         \caption{Best solution obtained using simulated annealing with 1000 shots.}
         \label{fig:SimAnsol1}
     \end{subfigure}
     \hfill
     \begin{subfigure}[t]{0.49\textwidth}
         \centering
         \includegraphics[width=\textwidth]{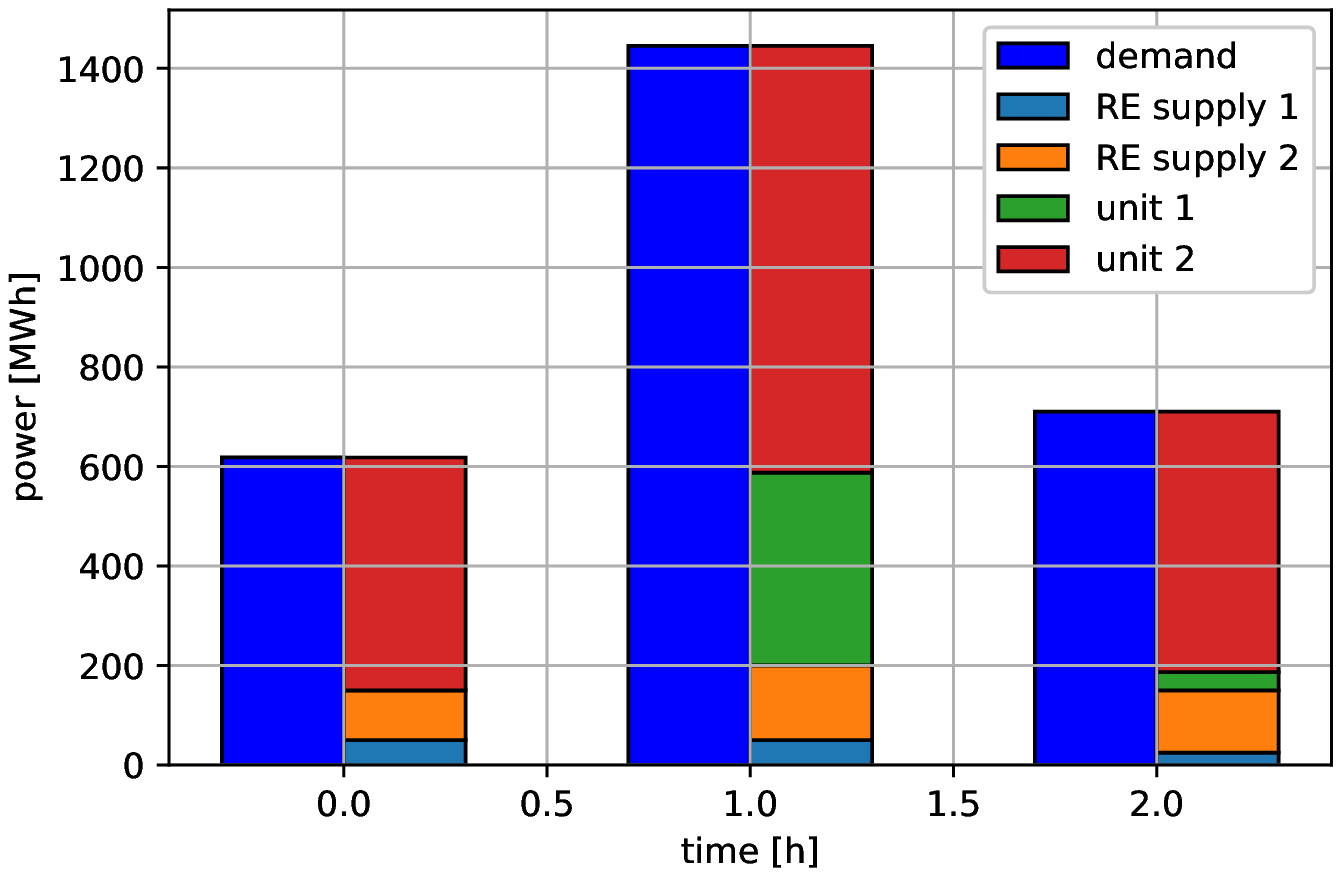}
        \caption{Best solution obtained using quantum annealing on D-Wave's Advantage 5.2 with 1000 shots}
        \label{fig:QuaAnsol1}
    \end{subfigure}
     \begin{subfigure}[t]{0.49\textwidth}
         \centering
         \includegraphics[width=\textwidth]{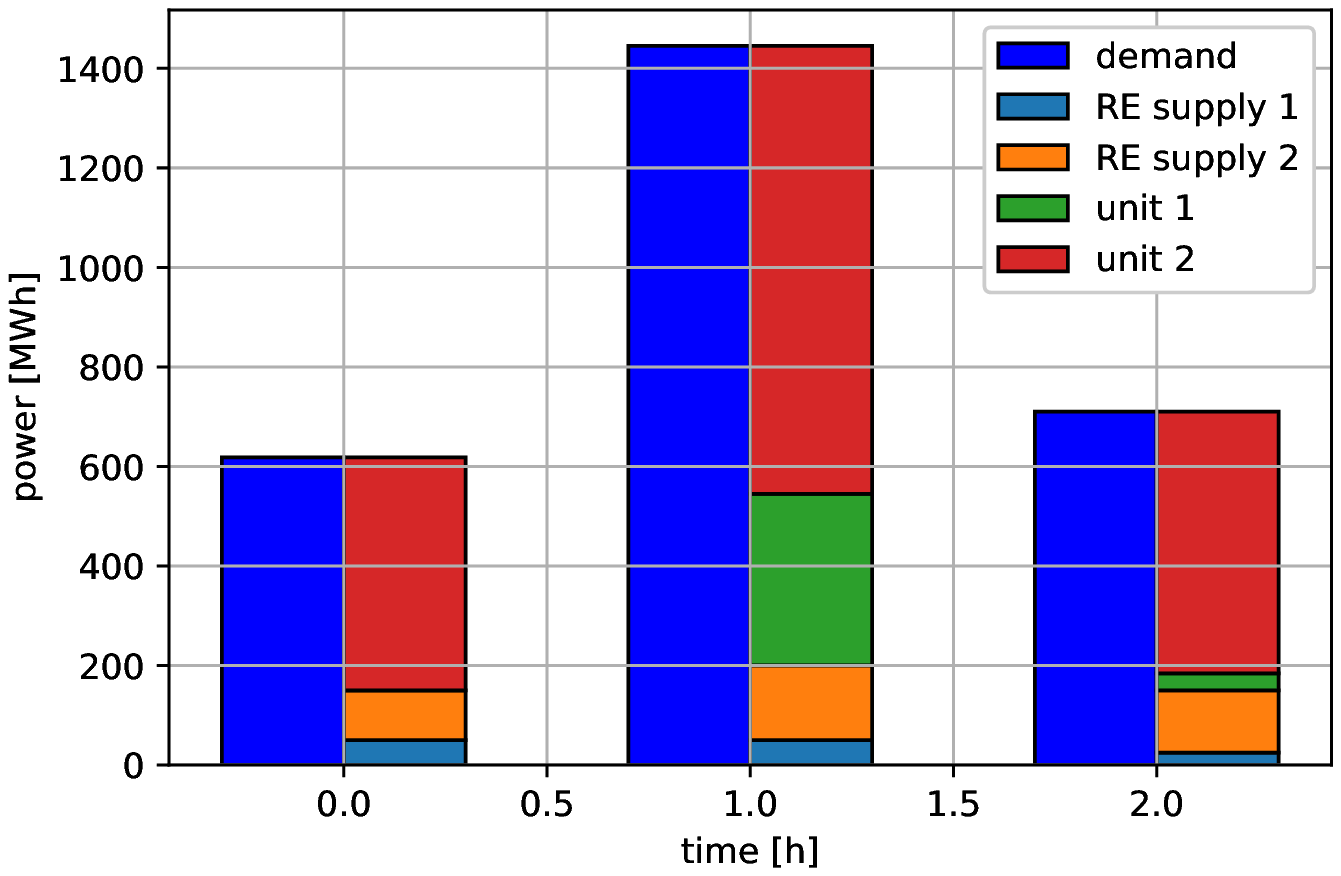}
         \caption{Optimal power unit schedule for the UCP specified in table~\ref{tab:unitparameters1} and table~\ref{tab:demand1}.}
         \label{fig:optsol1}
     \end{subfigure}
     \hfill
      \begin{subfigure}[t]{0.49\textwidth}
         \centering \raisebox{\dimexpr\ht\imagebox+0.15\height}{
            \includegraphics[width=\textwidth]{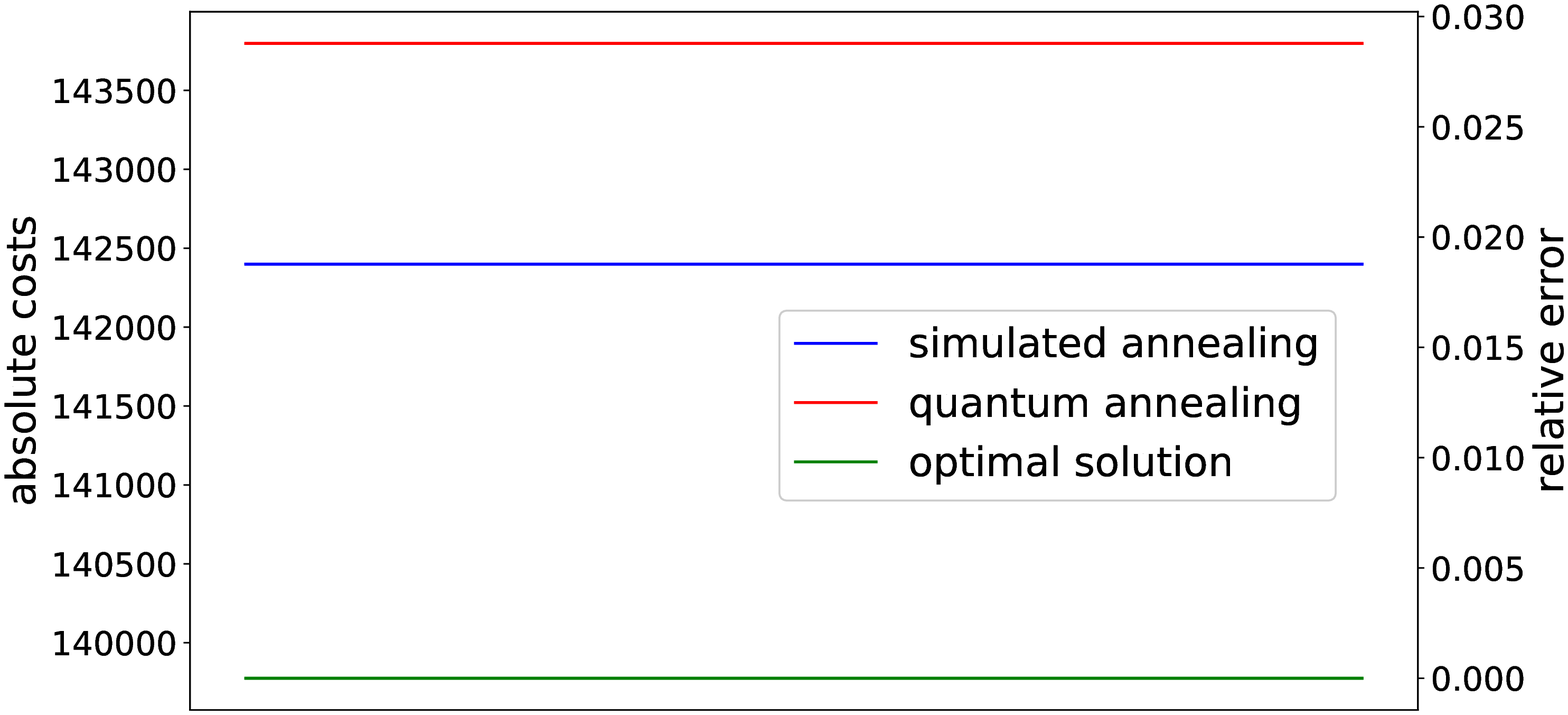}}
         \caption{Absolute costs and relative costs error of the solutions from simulated annealing and quantum annealing compared to costs of optimal solution.}
         \label{fig:comp1}
     \end{subfigure}
     \caption{Comparison of results from the two different solvers to the optimal solution. \ref{fig:SimAnsol1} - \ref{fig:optsol1} show the calculated power unit schedules and costs are compared in \ref{fig:comp1}.}
\end{figure}

These parameters fix the elements of the resulting $(72 \times 72)$-dimensional QUBO matrix, which is illustrated in figure~\ref{fig:qubo1} as a heatmap.
In this case, it is easy to find the optimal solution, which is shown in figure~\ref{fig:optsol1}. The minimum costs are given by $c_{\rm min}=139775$.

We now compare the results from a simulated annealing algorithm \footnote{D-Wave's simulated annealing sampler available in the neal-package.} with results from real quantum annealing hardware. We performed quantum annealing on the D-Wave Advantage system 5.2 containing 5760 qubits~\cite{dwaveadvantage}. After one run of simulated or quantum annealing, we subsequently performed a gradient descent method to find the nearest local minimum. Figure \ref{fig:comp1} compares the resulting costs from simulated annealing and quantum annealing to the optimal solution. It should be emphasized that the RE supply units that appear in table~\ref{tab:demand1} are not directly part of the optimization process, but contribute with a fixed amount of energy in each timestep which is previously set in the problem parameters.

It can be seen that all three solutions provide the same power unit schedule, but they assign slightly different power levels to the units which explains the small difference in the costs. Nevertheless, in this case, simulated annealing as well as quantum annealing find solutions of high quality, which both are approximately equal to the optimal power unit schedule.

\subsection{Benchmarking the formulation and solvers}

\subsubsection{Evaluating results}
In this section, we introduce an evaluation function that can be used as a measure for the quality of the model with respect to the solutions it produces. It enables us to compare the results from different solving methods. We introduce the probability $\nu$ for a solution to violate hard constraints. This values is a priori unknown. We estimate $\nu$ by performing 10 runs of one solver and divide the number of results with violated constraints by the total number of runs. We denote this quantity $\bar{\nu}$ and use it as an estimator for $\nu$. Furthermore, we define the critical probability $\nu_{\rm crit}$ of constraint violations that is tolerated. If we denote the minimal costs of the optimal solution as $c_{\rm min}$ and the costs obtained from one run of the algorithm as $c$, then we can define
\begin{equation*}
    f(\nu,c)=\begin{cases}
                \left(e^{-\frac{\nu}{\nu_{\rm crit}}}-\nu e^{-\nu_{\rm crit}^{-1}} \right)\cdot \left( 1- \frac{|c-c_{\rm min}|}{c_{\rm min}} \right)~& {\rm for}\; \nu_{\rm crit} \in (0,1] \\
                \Theta(-\nu)\cdot \left( 1- \frac{|c-c_{\rm min}|}{c_{\rm min}} \right)~& {\rm for}\;  \nu_{\rm crit}=0
              \end{cases}
\end{equation*}  
to be our evaluation function, where 
\begin{equation*}
    \Theta(x)=\begin{cases}
                0 \;{\rm for}\;  x<0 \\
                1 \;{\rm for}\;  x \geq 0
              \end{cases}
\end{equation*}  
is the Heaviside step-function. Note that in most cases $c_{\rm min}$ will be unknown. However, here we only consider cases where we know the exact optimal solution. The critical value $\nu_{\rm crit}$ should be chosen such that problem specific requirements are met. In our case, we choose a zero-tolerance policy and we set $\nu_{\rm crit}=0$, i.e. in the following we use 
$$
f(\bar{\nu},c)=\Theta(-\bar{\nu})\cdot \left( 1- \frac{|c-c_{\rm min}|}{c_{\rm min}} \right)
$$ 
for evaluating our results. When we calculate the value of $f(\bar{\nu},c)$ several times with different experiments, then we denote the average value of it by $\overline{f(\bar{\nu},c)}$.

\subsubsection{Simulated annealing vs. quantum annealing}

In the following, we present the results of comparing simulated annealing with quantum annealing for several examples. We introduce the following examples with different problem sizes:

\begin{table}[H]
    \captionsetup{width=0.87\linewidth,font={small},format=hang}
    \centering
    \caption{We define the number $N$ of conventional units and the number $T$ of timesteps for different example set sizes. The length $\dim{(x)}$ of the corresponding solution vector equals the number of rows and columns of the QUBO matrix for a resolution of $B=10$.}
    \begin{tabular}{c|c|c|c|c|c|c|c}
         example set & XXS & XS & S & M & L & XL & XXL  \\
         \hline
         [$N$, $T$] & [2,1] & [2,3] & [2,5] & [5, 24] & [50, 24] & [500, 24] & [5000, 24] \\
         \hline
         $\dim{(x)}$ & 24 &  72 &  120 & 1440 & 14400 & 144000 & 1440000 
    \end{tabular}
    \label{tab:benachmarksizes1}
\end{table}

Modeling very large power plant grids like the European electricity network, which contain thousands of power plants would lead to problems of similar size as the XXL example.
Due to the current size of the quantum annealers from D-Wave (Advantage system 5.2, Europe) the comparison is only feasible for our example sets XXS, XS and S with $B = 10$, i.e. a resolution of 10 bits.
\\

\begin{figure}[H]
\captionsetup{width=0.9\linewidth,font={small},format=hang}
    \centering
    \includegraphics[width=0.9\textwidth]{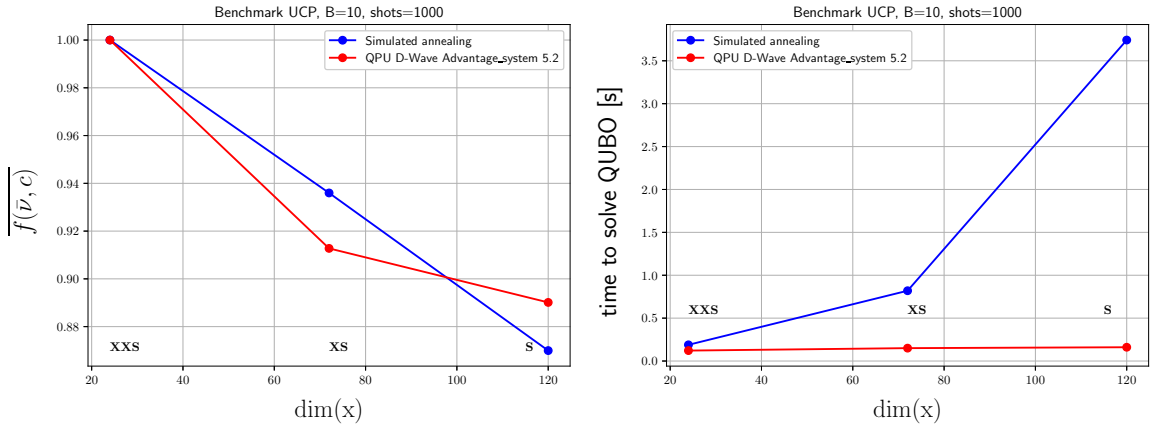}
    \caption{Solution quality $\overline{f(\overline{\nu},c)}$ (left) and time so solution in seconds (right) over the  QUBO size ${\rm dim}(x)$. The different example set sizes are marked.}
    \label{fig:solutionquality1}
\end{figure}

For the small problem set XXS with a QUBO size of  $24\times24$, both simulated annealing and the quantum annealer find the global solution. For bigger problem sets, the quality of the solutions is comparable but the runtimes differ vastly. For the problem set S, simulated annealing takes 3.6 seconds \footnote{We used a laptop with a 10 core CPU and the D-Wave package dwave-neal 0.5.9.} while the quantum annealer takes only 0.25 seconds. For simulated annealing, the pro-cessing time was measured using D-Wave’s neal package. For quantum annealing, the  "\texttt{qpu\_access\_time}" \footnote{D-Wave Operation and Timing: \url{https://docs.dwavesys.com/docs/latest/c_qpu_timing.html}} was taken, which does not consider pre-, post-processing, embedding time and queuing time. This gives us a hint that also for big problem sizes the time to an approximate solution increases only moderately.

\section{Relaxed QUBO with uncertainty}

\subsection{Elements with uncertainty}
In reality, it is not possible to predict the exact amount of future electricity demand or supply. Rather, the forecasting is subject to different kinds of uncertainty, especially in the exact energy consumption and the power supply from renewable supply units. However, we can account for these uncertainties by considering probability distributions for individual future scenarios. The task is then to find an optimal unit allocation strategy that exploits the stochastic character of the problem. Consequently, before continuing, we need to establish a stochastic framework.
\\ \hfill \break
Let us denote by $\mathcal{R}$ a set of renewable energy supply units with $R$ elements. In timestep $t$, for each renewable unit $r \in \mathcal{R}$ we introduce a discrete random variable $S^{\rm RE}_r(t)$ whose realizations $s^{i_r}_r(t)$ with $i_r\in \{1,2,\dots,n_r(t)\}$ 
 represent possible power supply stages of that unit. Moreover, a second discrete random variable for the demand $D(t)$, which can take values $\delta^j(t)$ for $j \in \{1,2,\dots,n_D(t)\}$ 
is used to describe the uncertainty in demand at time $t$. Having these variables, we can define the probability 
\begin{equation*}
p_r^{i_r}(t)=\Pr[S^{\rm RE}_r(t)=s^{i_r}_r(t)] 
\end{equation*}
for every power supply realization 
of unit $r \in \mathcal{R}$ and 
and the probability
\begin{equation*}
    p_D^j(t)=\Pr[D(t)=\delta^j(t)]
\end{equation*} 
for a specific demand at time $t$.
Now, a specific effective demand value is given by
 \begin{equation}\label{effectivedemand}
     d^{(j,i_1,\ldots,i_R)}(t)=\delta^j(t)-\sum_{r \in \mathcal{R}} s^{i_r}_r(t).
 \end{equation}
Since the effective demand is a function of two independent random variables $S^{\rm RE}_r$ and $D$, it also represents a random variable with the realizations of equation~\eqref{effectivedemand} to which the probabilities
 \begin{equation*}
     p_d^{(j,i_1,\ldots,i_R)}(t)=\Pr[d^{(j,i_1,\ldots,i_R)}(t)]=p_D^j(t) \cdot \prod_{r \in \mathcal{R}} p_r^{i_r}(t)
 \end{equation*}
 can be assigned. A scenario for the UCP is then defined by the collection of the realizations
$ d^{(j,i_1,\ldots,i_R)}(t)$
for $t\in\{1,\ldots, T\}$.
The probability for a specific scenario is then given by
\begin{equation*}
 \prod_{t=1}^T \prod_{r \in \mathcal{ R}} p_D^j(t)\cdot  p_r^{i_r}(t).
\end{equation*}
We will also need the expectation value for the effective demand:
\begin{equation*}
\Braket{d}(t)= \prod_{r \in \mathcal{R}} \prod_{i_r=1}^{n_r(t)} \prod_{j=1}^{n_D(t)} p_d^{(j,i_1,\ldots, i_R)}(t) \cdot d^{(j,i_1,\ldots, i_R)}(t).
\end{equation*}

Having defined these quantities, our aim is now to use them in a formalism to find an optimal power unit allocation such that we exploit the uncertainty contained in the probability distributions. We could simply look for the most probable scenario.
However, this strategy fails in the case of a uniform probability distribution. A similar approach that incorporates the stochastic aspects of the problem more profoundly is to work with the expectation values for the effective demand in every timestep and to optimize the expected scenario. However, this also does not exploit the stochastics and ignores the additional degrees of freedom that enter the problem through the underlying uncertainties. 

We propose a method that exploits the variety of possible effective demands in every timestep by introducing a relaxation of the optimization problem. Besides searching an optimal power unit allocation, we include the demand and renewable power supply in the optimization. We do this by allowing these quantities to take unique values $S^{\rm RE}_r(t)$ and $D(t)$ in all timesteps. The selection of these values is controlled via a minimization of
\begin{equation}\label{eq:compromisdemand}
\left( \sum_{t=1}^T \big[ d(t) - \Braket{d}(t) \big] \right)^2
\end{equation}
in addition to the cost minimization explained in section \ref{buildqubo}. Note that we use $d(t)$ here not as a random variable, but as a placeholder for the optimal realization that has to be found. The term \eqref{eq:compromisdemand} plays a similar role as a measure for the variance with respect to the expected scenario. This ensures that the power demand to be matched is close to the expectation values, but this formulation allows greater deviations from it, in favor of lower costs. We now show how to adapt the QUBO model such that this strategy can be implemented.

\subsection{Building multi-scenario QUBO's}
To take the minimization of the value~\eqref{eq:compromisdemand} into account, the  term
\begin{equation}\label{eq:compromisspenalty}
    P^{\rm var} \cdot \left( \sum_{t=1}^T \big[ d(t) - \Braket{d}(t) \big] \right)^2
\end{equation}
is added to the QUBO of equation~\eqref{eq:fullQUBO}.
Here, $P^{\rm var}$ is the penalty strength of the new term. Since we want the model to choose between different values for demand and renewable supply, we introduce more auxiliary variables to our solution vector. These contain information about whether a specific power stage is chosen or not. For simplicity, we choose a constant amount $n_R$ of possible power stages for all renewable units and $n_D$ demand values for every timestep. This leads to an overhead of $T\cdot\left( R \cdot n_R+n_D \right)$ variables. If we denote the new solution vector to this UCP including uncertainty as $x_{\rm relaxed}$, then we can attach the auxiliary variables to the end of the solution vector of equation~\eqref{eq:solvector} and we get
\begin{equation}\label{eq:stochsolvector}
    x_{\rm relaxed}=\left(x,x_{\rm RE}(1),x_{\rm RE}(2),\dots,x_{\rm RE}(T),e_{D}(1),e_{D}(2),e_{D}(T) \right).
\end{equation}
This new solution vector is of length 
$$
\dim{(x_{\rm relaxed})}=T\cdot \left(N(B+2)+R \cdot n_R+n_D\right)
$$
and $x_{\rm RE}(t)=(e_{{\rm RE}_1}(t),e_{{\rm RE}_2}(t),\dots,e_{{\rm RE}_R}(t))$ is a vector containing $R \cdot n_R$ elements for the power stages of the renewable energy suppliers. The unit length vectors $e_D(t)$ encode the chosen demand in a timestep and the unit length vectors $e_{{\rm RE}_r}(t)$ with $r\in \{1,\dots,R\}$ encode the power stage of RE-unit $r$ in timestep $t$. Consequently, with the vectors
$$
e_{{\rm RE}_{r;i}}=(0,\dots,\underbrace{1}_{i\text{th}},\dots,0)
$$ 
and $s_r(t)$, which holds the elements $s_r^1(t),\dots,s_r^{n_R}(t)$, the choice of power stage $i$ for the RE-unit $r$ in timestep $t$ is equivalent to evaluating $\langle e_{{\rm RE}_{r;i}}  | s_r(t)\rangle$. A specific unique effective demand is calculated via
\begin{equation}\label{eq:multiscenariodemand}
    d(t)= \Braket{e_D(t)|D(t)}  - \sum_{r \in \mathcal{R}} \Braket{e_{{\rm RE}_r}(t)|s_r(t)}, 
\end{equation}
where $D(t)$ contains the elements $\delta^1(t),\dots,\delta^{n_D}(t)$.

We need to add terms that ensure that we choose unique values for demand and the power stages of the renewables. These are 
\begin{equation}\label{eq:uniqueD}
    P^{\rm setD} \cdot \sum_{t=1}^{T} \bigg( |e_D(t)| -1 \bigg)^2\; {\rm and}
\end{equation}
\begin{equation}\label{eq:uniqueRE}
P^{\rm setRE} \cdot  \sum_{t=1}^{T}  \sum_{r \in \mathcal{R}} \bigg( |e_{{\rm RE}_r}(t)| -1 \bigg)^2
\end{equation}
with the individual penalty strengths $P^{\rm setD}$ and $P^{\rm setRE}$. Here,  
$|e_D(t)|$ and $|e_{{\rm RE}_r}(t)|$ are the sums of the binary entries of the vectors.

The new QUBO is given by
\begin{align*}
    \begin{split}
    Q_{\rm relaxed}(x)= ~Q(x) 
    +P^{\rm var}\left( \sum_{t=1}^T \big[ d(t) - \Braket{d}(t) \big] \right)^2  
    +P^{\rm setD} \cdot \sum_{t=1}^{T} \bigg( |e_D(t)| -1 \bigg)^2
    +P^{\rm setRE} \cdot  \sum_{t=1}^{T}  \sum_{r\in\mathcal{R}} \bigg( |e_{{\rm RE}_r}(t)| -1 \bigg)^2
    \end{split}
\end{align*}
Due to the structure of the solution vector $x_{\rm relaxed}$ from equation~\eqref{eq:stochsolvector}, the new QUBO matrix has the following structure:
\begin{equation*} Q_{\rm relaxed} = 
    \begin{pmatrix}
    Q & Q_{\rm couple} \\
    0 & Q_{\rm new} 
    \end{pmatrix}
\end{equation*}
The submatrix $Q_{\rm new}$ contains the corresponding matrix elements of~\eqref{eq:compromisspenalty}, \eqref{eq:uniqueD} and~\eqref{eq:uniqueRE}. The submatrix $Q_{\rm couple}$ ensures that we can choose between all possible scenarios and it originates from replacing the $d(t)$ value in equation~\eqref{eq:fullQUBO} with $d(t)$ from equation~\eqref{eq:multiscenariodemand}. As before, $Q$ is responsible for finding the best power unit schedule and contains the same information as \eqref{eq:fullQUBO}. Finally, we need to fix the numerical values of $P^{\rm setD}$, $P^{\rm setRE}$ and $P^{\rm var}$. Since we highly prioritize the uniqueness conditions, but we do not require a dominance of the $P^{\rm var}$ term, we empirically find the suitable values
$$    
P^{\rm setD}=P^{\rm setRE}=10^8 \cdot P^{\rm demand} \quad {\rm and} \quad
    P^{\rm var}=10^{-2} \cdot  P^{\rm demand}.
$$

\subsection{Example}\label{sec:example2}
We now come back to the example set XS, which was introduced in section \ref{example1}. We extend the example by considering uncertainty in the demand and in the production of the renewable units, as it can be seen in table~\ref{tab:demand2}.

\begin{figure}[H]
    \captionsetup{width=0.9\linewidth,font={small},format=hang}
    \centering
    \includegraphics[width=0.65\textwidth]{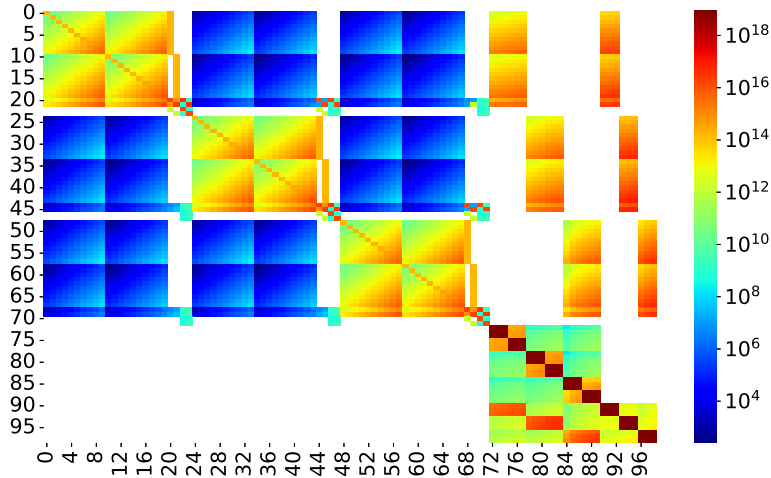}
    \caption{Heatmap of the $(99 \times 99)$-dimensional QUBO matrix for the UCP including uncertainty for the example set XS with resolution $B=10$. The parameters are specified in table~\ref{tab:unitparameters1} and table~\ref{tab:demand2}.}
    \label{fig:qubo2}
\end{figure}

\begin{table}[H]
    \captionsetup{width=0.9\linewidth,font={small},format=hang}
    \centering
    \caption{Probability distribution for power demand and renewable supply for the individual timesteps. This is an extension of the example set XS.}
    \begin{tabular}{c|c|c|c}
          \multirow[c]{2}{0.7cm}{\centering $t$} & \multirow[c]{2}{3.3cm}{\centering $D(t)$ \\ $\Pr{[D(t)]}$} & \multirow{2}{3.9cm}{\centering $S^{\rm RE}_1(t)$ \\ $\Pr{[S^{\rm RE}_1(t)]}$} & \multirow{2}{3.9cm}{\centering $S^{\rm RE}_2(t)$ \\ $\Pr{[S^{\rm RE}_2(t)]}$} \\  
          & & &\\
         \hline
          \multirow[c]{2}{0.7cm}{\centering 1} & \multirow[c]{2}{3.3cm}{\centering $\{518,618,718\}$ \\ $\{0.1,0.8,0.1\}$} & \multirow[c]{2}{3.9cm}{\centering $\{40,50,60\}$ \\ $\{0.1,0.8,0.1\}$} & \multirow{2}{3.9cm}{\centering $\{80,100,120\}$ \\ $\{0.1,0.8,0.1\}$}   \\
          & & & \\
          \hline 
          \multirow[c]{2}{0.7cm}{\centering 2} & \multirow[c]{2}{3.3cm}{\centering $\{1145,1145,1745\}$ \\ $\{0.15,0.7,0.15\}$} & \multirow{2}{3.9cm}{\centering $\{30,50,70\}$ \\ $\{0.15,0.7,0.15\}$}  & \multirow{2}{3.9cm}{\centering $\{120,150,180\}$ \\ $\{0.15,0.7,0.15\}$}  \\
          & & & \\
          \hline
          \multirow[c]{2}{0.7cm}{\centering 3} & \multirow[c]{2}{3.3cm}{\centering $\{310,710,1110\}$ \\ $\{0.2,0.6,0.2\}$} & \multirow[c]{2}{3.9cm}{\centering $\{5,25,45\}$\\ $\{0.2,0.6,0.2\}$}  & \multirow[c]{2}{3.9cm}{\centering $\{80,125,170\}$ \\ $\{0.2,0.6,0.2\}$} \\
          & & & 
    \end{tabular}
    \label{tab:demand2}
\end{table}
\hfill \break
A heatmap of the QUBO matrix that is built from the parameters from table~\ref{tab:unitparameters1} and the probability distributions from table~\ref{tab:demand2} is shown in figure~\ref{fig:qubo2}.

\begin{figure}[H]
\captionsetup{width=0.87\linewidth,font={small},format=hang}
     \centering
     \begin{subfigure}[t]{0.48\textwidth}
         \centering
         \includegraphics[width=\textwidth]{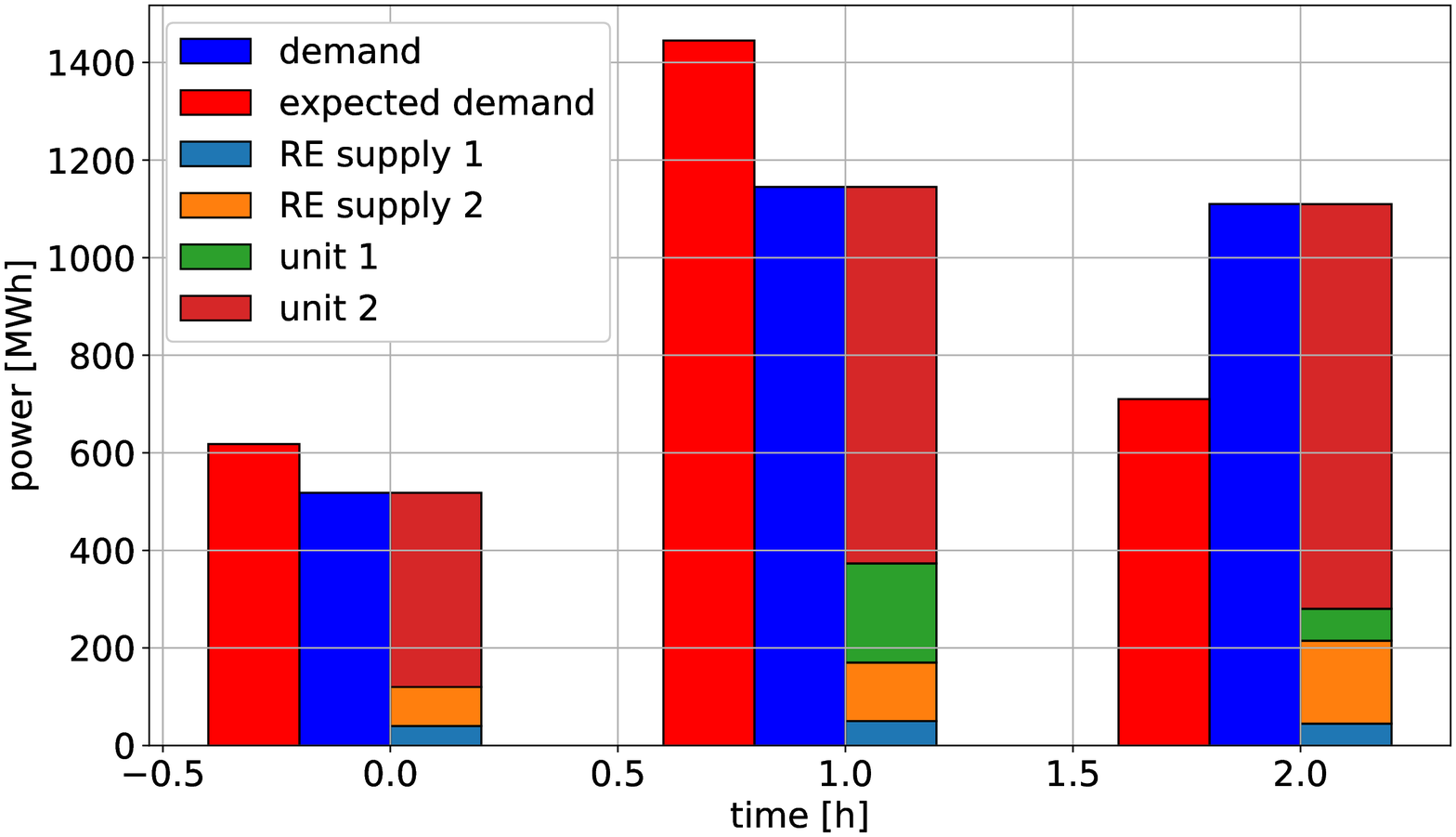}
         \caption{Best solution obtained from one run of simulated annealing (cost=137447).}
         \label{fig:SimAnsol2}
     \end{subfigure}
     \hfill
     \begin{subfigure}[t]{0.48\textwidth}
         \centering
         \includegraphics[width=\textwidth]{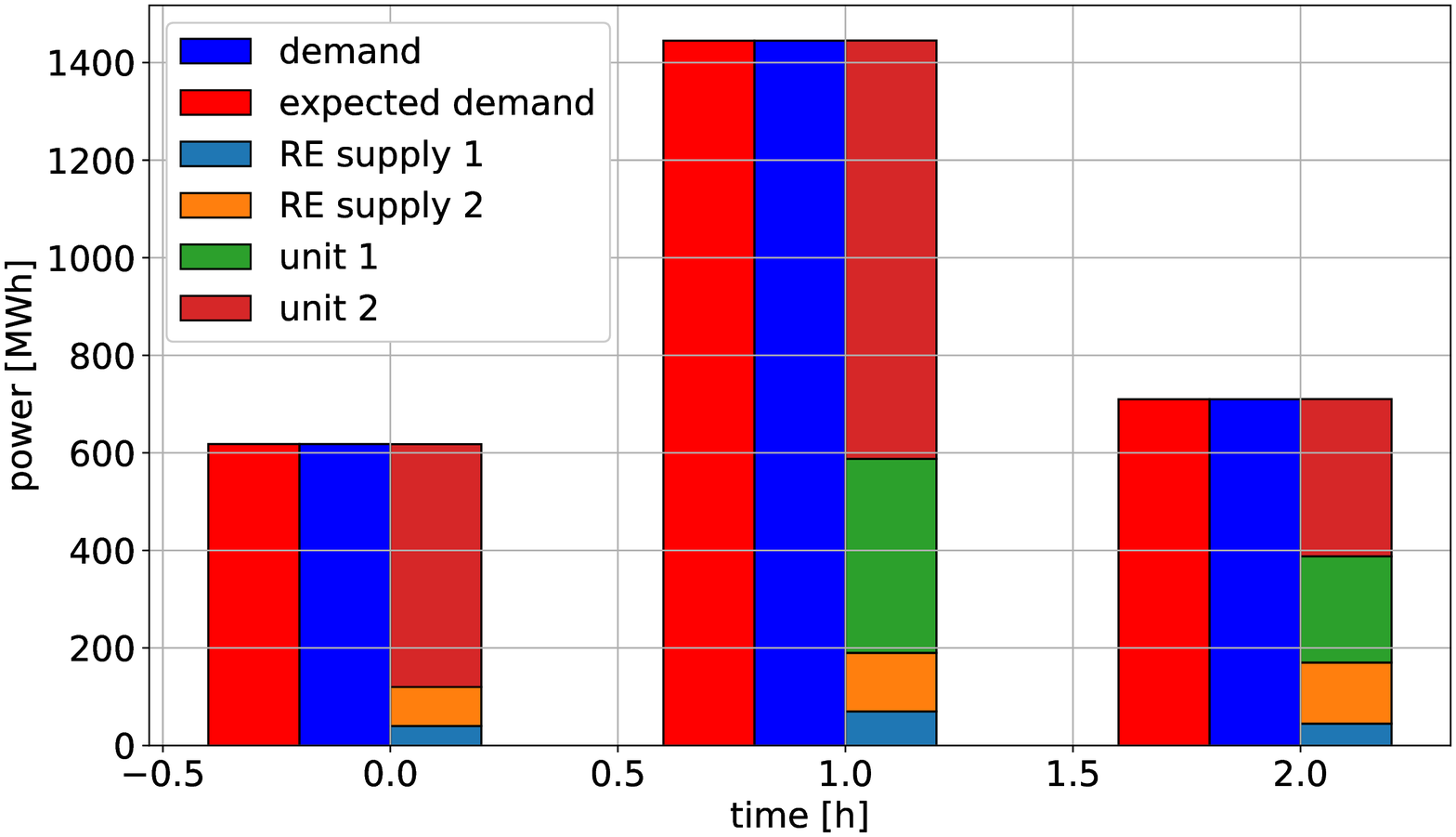}
        \caption{Best solution obtained from one quantum annealing run on D-Wave Advantage 5.2 (cost=151944).}
        \label{fig:QuaAnsol2}
    \end{subfigure}
     \begin{subfigure}[t]{0.48\textwidth}
         \centering
         \includegraphics[width=\textwidth]{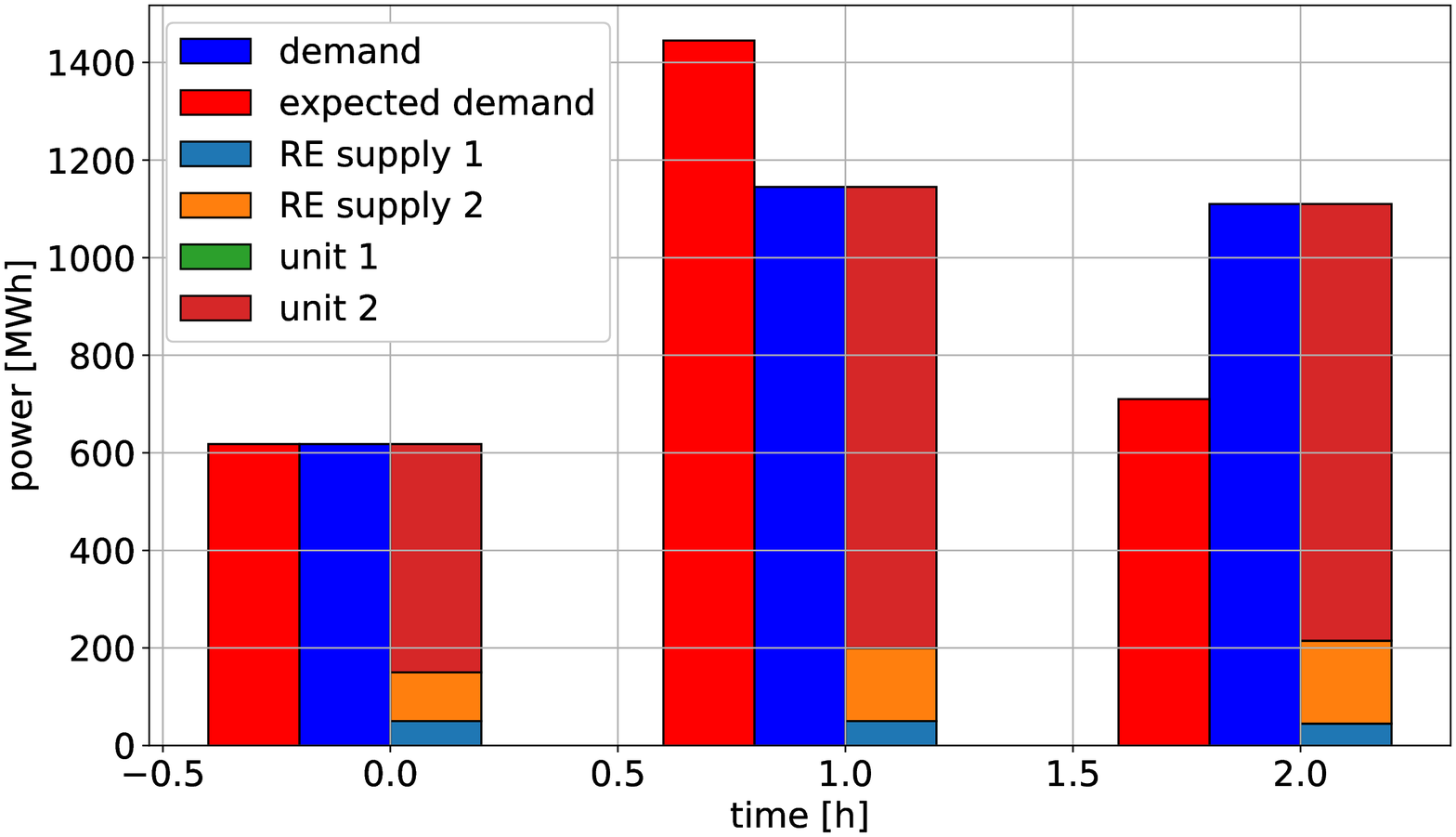}
         \caption{Optimal power unit schedule for the UCP specified in table~\ref{tab:unitparameters1} and table~\ref{tab:demand1} (cost=106950).}
         \label{fig:optsol2}
     \end{subfigure}
     \hfill
      \begin{subfigure}[t]{0.48\textwidth}
         \centering\centering \raisebox{\dimexpr\ht\imagebox+0.15\height}{
            \includegraphics[width=\textwidth]{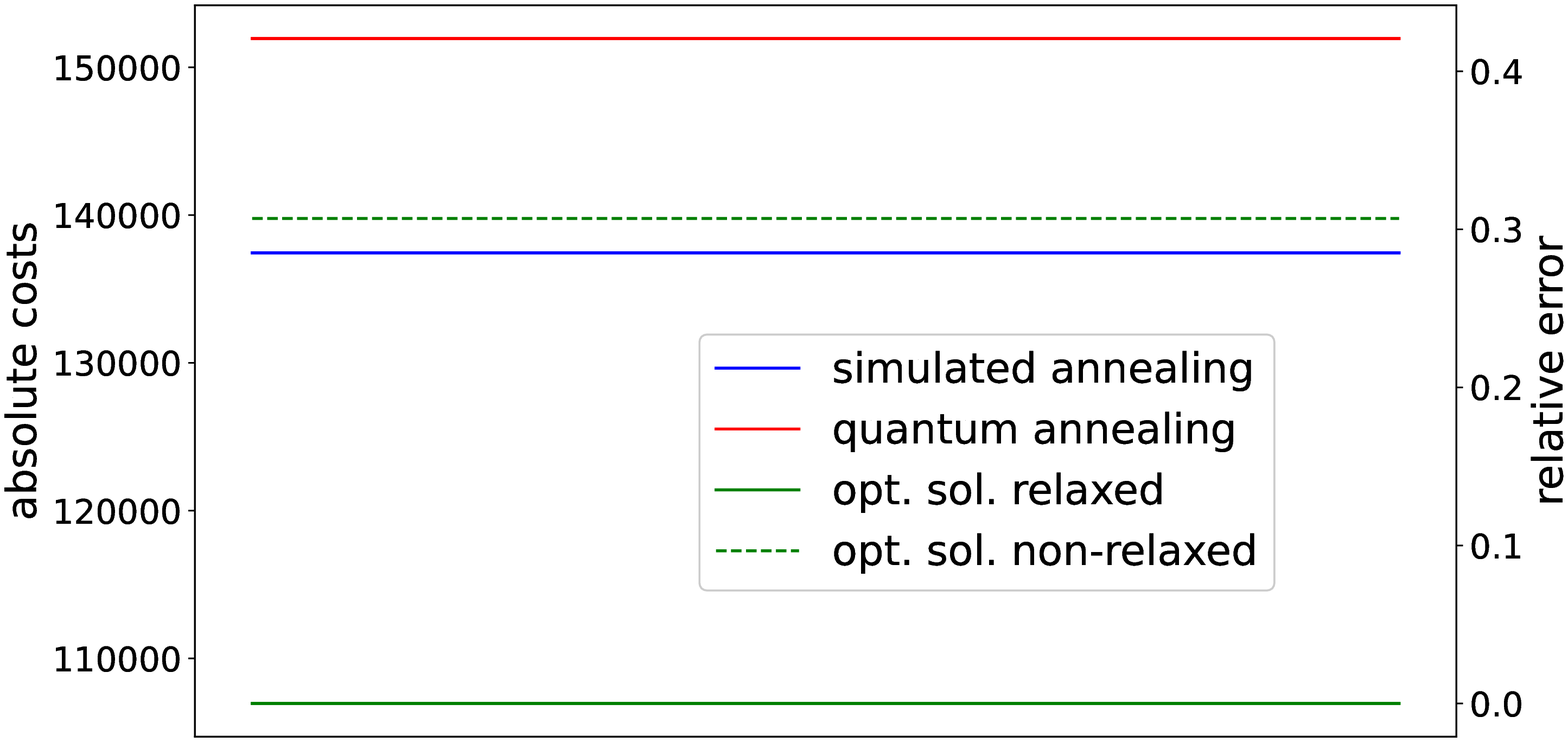}}
         \caption{Absolute costs and relative error of the solutions from simulated annealing and quantum annealing, compared to costs of optimal solution of the UCP and the relaxed UCP from section~\ref{example1}.}
         \label{fig:comp2}
     \end{subfigure}
     \caption{Comparison of results from the two different solvers to the optimal solution. Figures~\ref{fig:SimAnsol2} - \ref{fig:optsol2} show the calculated power unit schedules and costs are compared in figure~\ref{fig:comp2}.}
\end{figure}

\hfill \break
Again, we compare the results from simulated annealing (figure~\ref{fig:SimAnsol2}) and quantum annealing (figure~\ref{fig:QuaAnsol2}) to the optimal solution, shown in figure~\ref{fig:optsol2}.

\hfill \break
\vspace{-0.5cm}
\begin{figure}[H]
\captionsetup{width=0.87\linewidth,font={small},format=hang}
     \centering
     \begin{subfigure}[t]{0.48\textwidth}
         \centering
         \includegraphics[width=\textwidth]{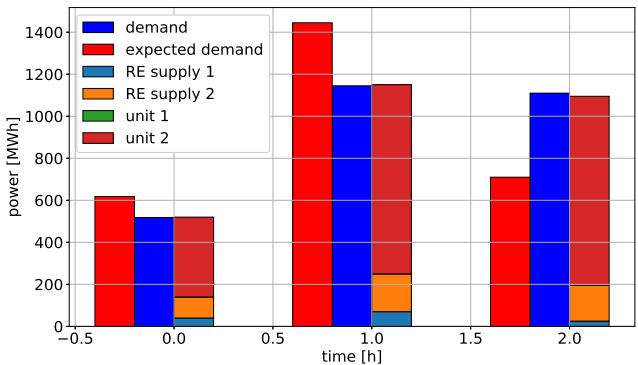}
         \caption{Best solution obtained from one run of simulated annealing (relative cost error=2.29 \%).}
         \label{fig:SimAnsol2B4}
     \end{subfigure}
     \hfill
     \begin{subfigure}[t]{0.48\textwidth}
         \centering
         \includegraphics[width=\textwidth]{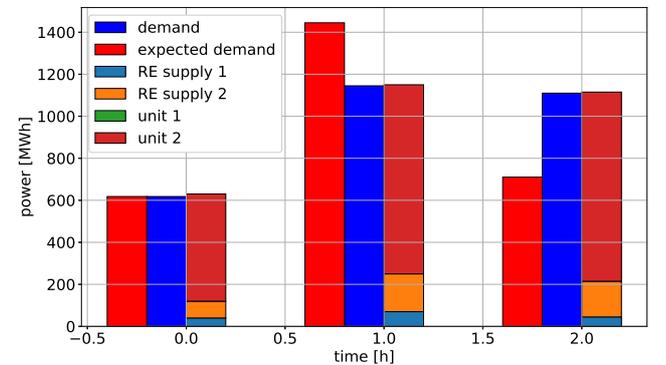}
        \caption{Best solution obtained from one run of quantum annealing (relative cost error=2.27 \%).}
        \label{fig:QuaAnsol2B4}
    \end{subfigure}
    \label{fig:compB4}
    \caption{Comparison of results from the two different solvers for a solution of $B=4$. Both solvers find the optimal power unit schedule.}
\end{figure}

The difference in the quality of the solutions of simulated annealing and quantum annealing is bigger than in the example without uncertainty due to the increased size of the QUBO.
However, it can be seen that simulated and quantum annealing find comparable solutions close but not exact to the optimal solution.
As a further example for the results of an optimization, figures~\ref{fig:SimAnsol2B4} and~\ref{fig:QuaAnsol2B4} show the solutions for $B=4$.

It can be seen that in this case quantum annealing is even able to beat simulated annealing. However, a smaller value for the resolution $B$ comes in general with a decreased resolution in demand matching.

\subsection{Benchmarking}
We now compare the results of simulated and quantum annealing for the UCP  with uncertainty. We introduce several examples with different problem sizes as shown in table~\ref{tab:benachmarksizes2}. The complexity of the formulation scales with the size 
\begin{equation*}
    T \cdot (n \cdot (B+2) + R \cdot n_r + n_D)
\end{equation*}
of the solution vector. 
\begin{table}[H]
    \centering
    \captionsetup{width=0.87\linewidth,font={small},format=hang}
    \caption{We consider $N$ units and $R$ renewable units in different example set sizes. Further, we have the number $n_R$ of power stages of the RE units, the number $n_D$ of power stages of the demand and $T$ timesteps. The resulting length of the solution vector equals the number of rows and columns of the corresponding QUBO matrix for a resolution with $B=10$.}
    \begin{tabular*}{0.87\textwidth}{@{\extracolsep{\fill} }c|c|c|c|c}
         example set &  XXS &  XS & S & M  \\
         \hline
         [$N$,$R$,$n_R$,$n_D$,$T$] & [2,2,2,2,1] &[2,2,2,2,3] & [2,2,3,3,5] & [5,3,5,5,24]  \\
         \hline
         ${\rm dim}(x)$ & 33 &  99 & 165 & 1920 \\
         \hline \hline
           example set & L & XL & \multicolumn{2}{c}{XXL} \\
         \hline
          [$N$,$R$,$n_R$,$n_D$,$T$] & [50,50,5,5,24] & [500,100,20,20,24] & \multicolumn{2}{c}{[5000,1000,10,10,24] }\\
         \hline
         ${\rm dim}(x)$ & 9000 & 168240 & \multicolumn{2}{c}{1680240}
    \end{tabular*}
    \label{tab:benachmarksizes2}
\end{table}

\begin{figure}[H]
\captionsetup{width=0.9\linewidth,font={small},format=hang}
    \centering
    \includegraphics[width=1.0\textwidth]{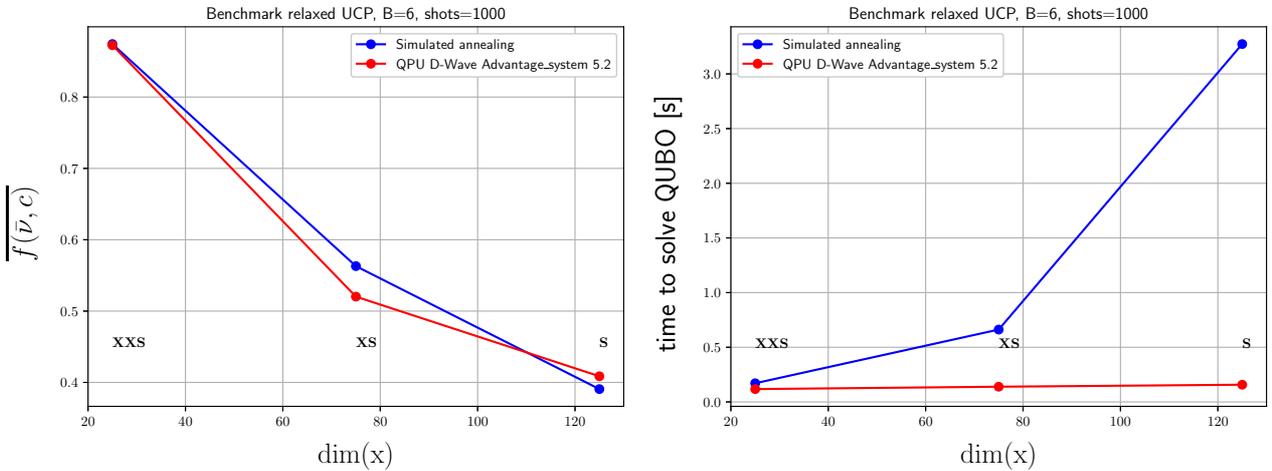}
    \caption{Solution quality $\overline{ f(\overline{\nu},c)}$ (left) and time to solution in seconds (right) over the QUBO size ${\rm dim}(x)$. The different example set sizes are marked.}
    \label{fig:solutionquality2}
\end{figure}

The comparison of simulated and quantum annealing using the D-Wave advantage 5.2 system is
shown in figure~\ref{fig:solutionquality2}.

It can be seen that the quantum annealer achieves a slightly better solution quality for bigger problem sets, e.g. we have a 3\% points improvement with example set S. In terms of runtime, the difference is much larger. For example set~S, the quantum annealer only requires 0.16 seconds, which is more than $14$ times faster than the simulated annealing method\footnote{We used a laptop with a 10 core CPU and the D-Wave package dwave-neal 0.5.9.}. For simulated annealing, the processing time was measured using D-Wave's neal package. For quantum annealing, the "\texttt{"qpu\_access\_time}" was taken.

\section{Discussion and outlook}
The hardness of the UCP suggests to reformulate the problem in such a way that advanced non-classical methods can provide shorter optimization times for finding solutions with a high quality. We showed that such a reformulation can be achieved by using QUBO's, which can be solved by Ising machines leveraging quantum computing techniques.

The complexity of the problem correlates with the size of the corresponding QUBO matrix. Including more technical constraints may lead to a larger overhead, i.e., introducing additional auxiliary variables like $x^s$ or $x^{1/0}$ becomes necessary. Regardless of the size of the QUBO, it is possible to construct a correct mathematical model in which low energy states represent feasible solutions.

The problem comes with the sampling of states corresponding to the desired solution. Algorithms like simulated annealing fail because the search space is too big. The large number of states make it unlikely for an algorithm that runs for a limited amount of time to find the global minimum or solutions with a sufficiently low energy. Here, quantum annealing provides a possible solution, at least at the theoretical level. In the optimal case, the system would always end up in the global minimum. However, as it can be seen from our results, current state-of-the-art machines are far away from providing a real advantage over simulated annealing. While good solutions are found for small problem sizes, the hardware is still several orders of magnitude too small for real-world problems.

For example, a model of the German electricity market would require to include roughly 2000 conventional units alone~\cite{Bundesnetzagentur}. The number of renewable power supply stations increases steadily. Therefore, a model including all power stations individually seems to be unrealistic for several reasons, including a lack of computational resources and sufficiently large quantum annealing machines. However, a model of local networks with a smaller number of energy supply units might be possible.

Besides the shortcomings of the current Ising machines, difficulties arise when the QUBO matrix is passed to the hardware, which needs to map the problem to the QPU topology. This mapping is called minor embedding and requires time, which has to be added to the actual solving time of the machine. An increasing problem size also leads to longer embedding times, which, at some point, start to exceed the actual solving time. A pre-defined embedding routine might fix this issue,
but this would strongly depend on the yet unknown topology of bigger QPU's.

We conclude that our quantum approach to the UCP is a promising alternative to classical solutions under the assumptions that future annealing machines will eventually reach appropriate levels of size and accuracy. The ongoing ambitious developments in the field give hope for a future generation of quantum annealers, which may actually be able to solve such hard problems like the UCP efficiently.

Besides this technical point of view, we also would  like to discuss the underlying principle that is used in our construction. Searching for a solution that lies in the neighborhood of a mean value scenario is a first approach to treat uncertainties. Since we choose to  match only the accumulated mean demand over time, it happens that sometimes the solution underestimates the power demand in one timestep and overestimates it in the next, as it is the case for the example in section \ref{sec:example2}. Such a property of the solution would require elements in the power net that compensate such demand dips and tips. Energy storage systems provide such elements and can act as a damping element for fluctuations in the net.
A more advanced modeling should account for these storage systems and integrate them into the UCP.

Another issue concerns the principle of the approach we presented here. If one deals with probability distributions that have multiple maxima, which belong to quite separated scenarios, then this approach leads to insufficient solutions. For example, consider a solar park that is located in a geographical position such that a passing cloud front in an otherwise clear sky may or may not block the sun from shining on the solar panels over a period of time. In the most extreme case, one can then assign a 50\% chance of getting no power and a 50\% of getting the full supply, since the exact path that the cloud takes is hard to predict. Using our approach, we would find a solution near a scenario where 50 percent of the full solar power is supplied. Therefore, the situation that actually occurs could mean that we would have to change the calculated power unit schedule to compensate for a rather big demand deficit or surplus. 
This means that we have costs that were not considered in the original computation of a solution. 

To avoid this kind of problem, we suggest to modify our approach in such a way that we consider weighted costs for different scenarios, where the weight is determined by the costs it would take to change the power unit schedule in a worst case scenario. We then try to find the scenario that balances low operating costs and low schedule changing costs. 

As a final note, this work can also be seen as the foundation to benchmark further quantum algorithms (e.g. VQE and QAOA) that can also solve QUBO problems.

\section{Acknowledgments}
This work is supported by the Federal Ministry for Economics and Climate Action through the project ‘EnerQuant’ (Project-ID 03EI1025B).

\bibliographystyle{ieeetr}
\bibliography{ref}

\end{document}